# Observation of a Fifth of the Electron Charge

M. Reznikov[a], R. de Picciotto[b], T. G. Griffiths, M. Heiblum, and V. Umansky

*Braun Center for Submicron Research, Dept of Cond. Matter Physics,*

*Weizmann Institute of Science, Rehovot 76100, Israel*

[a] *Currently at Physics Department, Technion, Haifa 32000, Israel*

[b] *Currently at Bell-Labs, Lucent Technologies, Murray Hill, NJ, 07974, US*

The *fractional quantum Hall* (FQH) effect is a phenomenon observed in the conduction properties of a two-dimensional electron gas subjected to a strong perpendicular magnetic field. The effect results from the strong interaction among electrons, brought about by the high magnetic field, giving rise to *fractionally charged* quasi-particles that carry the current[1-5]. Here we report the observation of quasi-particles with a charge $q=e/5$ detected by shot noise measurements in the 2/5 conducting channel (filling factor 2/5). This is in agreement with previous measurements that showed that the current in the lower, *1/3*, channel is carried by quasiparticles with a charge $q=e/3$. These results demonstrate that the actual fraction of the charge can be different from the filling factor. Moreover, we show that there is no strong interaction between the channels, which can be considered as conducting the current independently.

The fractional quantum Hall (FQH) effect[1] is a manifestation of the prominent and unique effects resulting from the Coulomb interactions among the electrons in a two-dimensional electron gas (2DEG) under the influence of a strong magnetic field. Laughlin's seminal explanation of the FQH effect[1,2,3] involved the emergence of new, fractionally charged, quasi-particles. Recently, shot noise measurements[4,5] confirmed the existence of these quasi-particles in the FQH regime. Shot noise, resulting from the granular nature of the particles, is proportional to the charge of the current-carriers, in this case quasi-particles[4,5]. In those experiments a quantum point contact (QPC), embedded in a 2DEG, was used as an electronic 'beam splitter'. Its purpose was to partly reflect the incoming current and lead to partition of carriers and hence to shot noise. An applied magnetic field corresponding to a fractional filling, in the bulk far from the QPC, $v_B=1/3$ in Ref. 4 or $2/3$ in Ref. 5 was employed. Charge was deduced via the generalized equation for shot noise (the classical, simplified, version is the Schottky one: $S=2qI_B$, with $S$ the spectral density of current fluctuations, $I_B$ the reflected current, and $q$ the charge of the current-carrying particle). For small reflection by the QPC (small $I_B$) the quasi-particle's charge was found to be $e/3$ (Ref. 4 & 5); as predicted theoretically[6,7,8]. The theories were based on the chiral Luttinger liquid model. However, for other, more general, filling factors (such as $v=2/5$), such calculations become exceedingly complicated. Still, one can gain insight into the characteristics of the expected shot noise in such cases by considering the Composite Fermion (CF) model[9].

In the FQH regime the current is carried by quasi-particles with a charge $q=e/(2pn+1)$, with $e$ the electronic charge. The fractional filling factor $v=p/(2pn+1)$ determines the conductance of the sample $g=vg_0$ with $g_0=e^2/h$ being the quantum conductance. Within the CF model, fractional filling factors for electrons of the form $v=p/(2pn+1)$ are identified as integer filling factors of Composite Fermions, $v_{CF}=p$. Each CF is composed of a single electron with *2n* quanta of magnetic flux attached to it (each flux quantum is $\phi_0=h/e$). Here we deal with the simplest family of CFs, when only two flux quanta are attached to each electron (*n=1*). Filling factor *v=1/3* then corresponds to the simplest fraction *p=1*, while *v=2/5* corresponds to *p=2*. The effective magnetic field sensed by the CFs is $B-2n_s \cdot h/e$, with $n_s$ the density of the 2DEG, where the magnetic field attached to the CF themselves is subtracted. Under this, weaker, effective magnetic field these CFs are usually considered as non-interacting particles (though this justification might not be solid). Shot noise, induced by the QPC, is thus recognized as a partition noise - associated with partial reflection of CFs in integer Landau levels (to be named CF channels). This scenario is similar to the one taking place at a QPC reflector at zero magnetic field[10,11,12,13]. Hence, the shot noise in the weakly reflected channel *p* (filling factor *p/(2p+1)*) is expected to correspond to quasi-particles with charge *q=e/(2p+1)*.[14]

It is important to examine the noise properties at fractional filling factors in which the value of the charge in units of *e* is expected to differ from the filling factor. Observation of a charge *e/(2p+1)* for *p>1* will remove a possible ambiguity[15,16] in the interpretation of the experimental data. Namely, it will prove that our charge measurement is not simply a different way to measure the conductance (or the filling factor). An obvious experimental target is the filling factor *v=2/5*, where the charge of the current-carrying quasi-particles is expected to be *(1/5)e* while the conductance is *(2/5)$g_0$*.

The contribution of each 1D channel without magnetic field is known to be equal to the quantum conductance $g_0=e^2/h$. However, the conductance of each CF channel has a smaller, channel dependent, value $\delta g_p$. For example, the contribution of the 1st CF channel is $\delta g_1=g_0/3$ while that of the 2nd channel is $\delta g_2 = (\frac{2}{5}-\frac{1}{3})\cdot g_0$. Hence, when the two lowest CF channels are fully transmitted the higher channel carries a small portion (1/6) of the total current. Taking into account the smaller charge of the corresponding quasi-particle, $e/5$, a very weak shot noise signal is expected. This makes the measurement of the noise in the 2/5 channel exceedingly difficult.

Our experiment was performed with a setup similar to the one used in Refs. 4 & 17. Noise was measured within a bandwidth of ~*30 kHz* about a central frequency of *1.68 MHz*. This frequency was chosen by tuning an LRC resonant circuit, with C the capacitance of the coaxial cables. By choosing a two-fold lower frequency compared to the one used previously[4] we reduced the spurious noise contribution of our preamplifier from *1.1x10$^{-28}$ A$^2$/Hz* (at a central frequency of *4 MHz*) to *3.8x10$^{-29}$ A$^2$/Hz* in the present frequency range. The preamplifier, which operates at 4.2 K, was manufactured from transistors fabricated on GaAs-AlGaAs wafers grown in our own MBE system. Measurements were performed in a dilution refrigerator at an electron temperature *T=85 mK*. This temperature was deduced from comparing the thermal noise[4] without an external current through the QPC to the Johnson-Nyquist formula *S=4k$_B$Tg* via *T=($\delta$S/$\delta$g)/4k$_B$*, with *g* the total conductance. Extrapolating this linear dependence to zero conductance (of the QPC) gives the contribution of our preamplifier to the total noise.

Our sample was based on a 2DEG embedded in a GaAs-AlGaAs heterostructure. The 2DEG had a low temperature carrier concentration $n_s$=*1.15x10$^{11}$ cm$^{-2}$* and mobility *4.2x10$^6$ cm$^2$/Vs*. A perpendicular magnetic field of some *12 Tesla* led to a bulk filling factor 2/5. The QPC was formed by evaporating two metallic gates onto the surface of the heterostructure (see lower inset in Fig. 1). By applying bias to the gates with respect to the 2DEG we controlled the transmission of the incoming current through the small opening of the QPC. The measured two-terminal conductance of two different runs, at bulk filling factors $v=2/5$ and $v\sim1/2$, are shown in Fig.1. Note that for zero voltage on the gates of the QPC the value of the conductance, in both cases, was lower than the bulk value ($g<v\cdot g_0$). This is a result of unintentional reflection from the QPC even when it is not biased. Indeed, when a small positive gate voltage was applied to the QPC the conductance increased and reached its bulk value. Application of a negative gate voltage recovered the 2/5 and 1/3 states (see Fig. 1 and upper inset), as expected.

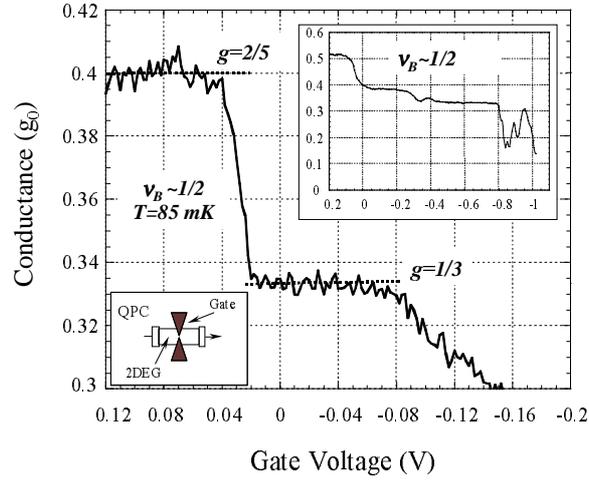

**Figure 1.** Two-terminal conductance (in units of the quantum conductance) against applied voltage to the gates of the Quantum Point Contact (QPC). Clear plateaus of $(2/5)g_0$ and $(1/3)g_0$ are seen when bulk filling factor is 2/5 ($B=12\ T$). **Top Inset:** A similar two terminal conductance plot for a bulk filling factor of 1/2 ($B=10.5\ T$). **Bottom Inset:** Schematic of the QPC in 2DEG. Applying voltage to the two gates restricts current flow in the constriction.

In the absence of an exact model for shot noise at $v=2/5$ we compare our results with the partition noise one expects from non-interacting quasi-particles. In other words we assume that the scattering events of the quasi-particles at the QPC are *independent*. This assumption was quite successful in analyzing the noise results for quasi-particles with charge $e/3$ in a single CF channel (Refs. 4 & 5). At zero temperature, one would expect a zero frequency spectral density of current fluctuations, $S$, given by:

$$S_{T=0} = 2qV\delta g_p t_p (1-t_p) \quad , \tag{1}$$

where $V$ is the applied bias voltage across the QPC, $\delta g_p$ is the contribution of the $p$'th CF channel to the total conductance, $t_p$ is the transmission coefficient of this channel and $q$ is the charge of each quasi-particle[13]. For example, for $p=2$, $\delta g_2 = g - g_0/3$ and $t_2 = \frac{g/g_0 - 1/3}{2/5 - 1/3}$.

A more subtle issue is the noise expected at a finite temperature $T$. This noise does not vanish at zero applied voltage but approaches the Johnson-Nyquist formula: $S = 4k_B T g$, in accordance with the fluctuation-dissipation theorem. When a bias voltage is applied, the noise is expected to increase smoothly with increasing $V$, approaching the linear behavior predicted by Eq. (1) at an applied voltage greater than $V_T \sim 2k_B T/q$. The detailed way in which this comes about in a FQH system is not known analytically (numerical calculations[18] exist for $v=1/3$ but contain parameters which are difficult to obtain experimentally). For that reason we adopted a form which was derived for non-interacting Fermions that interpolates between these two limits[19], as was also done, successfully, for single channel CF in Refs. 4 & 17:

$$S = 2qV\delta g_p t_p (1-t_p) \left[ \operatorname{ctanh}(\frac{qV}{2k_B T}) - \frac{2k_B T}{qV} \right] + 4k_B T g \quad . \tag{2}$$

We start with noise measurements on the conductance plateaus. The inset in Fig. 2 shows these results on both, the 2/5 ($t_2=1$, $t_1=1$) and 1/3 ($t_2=0$, $t_1=1$) plateaus. While in the first case two CF channels are fully transmitted in the latter the second one is fully reflected. We find no excess noise (above the thermal noise) within the accuracy of the measurement. This shows that the impinging current in both cases is noiseless as one would expect for two *separate* CF channels.

We come to this point later again. As the transmission of either channel is lowered, to be between zero and one, partitioning sets in and excess noise is observed as described below.

To tie the present results with previous experiments we start with measurements of noise generated by partly reflecting the first CF channel (the 1/3 channel). This time we pick a larger bulk filling factor $v_B \sim 1/2$. This is done in order to validate the previous measurements and verify that only the partitioning of the corresponding channel by the QPC is important in the determination of the charge of the quasi-particle. Figure 2 shows the measured noise for a QPC fully reflecting the second CF channel and only weakly reflecting the first CF channel with $t_1=0.9$, $t_2=0$. The data, after calibration and subtraction of the amplifier's contribution, agrees very well with the predicted shot noise with $q=e/3$ and $T=85$ $mK$ substituted in Eq. (2); in excellent agreement with Refs. 4 & 5.

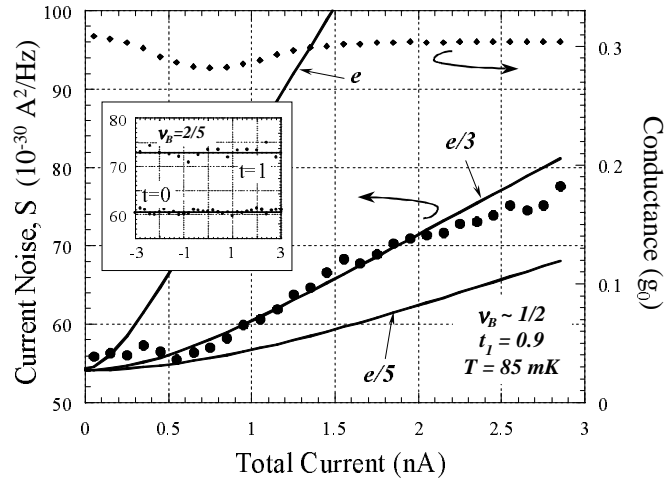

**Figure 2.** Measured spectral density of current fluctuations for $t_1=0.9$, $t_2=0$ (solid circles, left scale). The result agrees well with Eq. (2) and a quasi-particle charge $e/3$. For comparison the expected noise curves for quasi-particle charge $e/5$ and $e$ are shown. In the **Inset** the noise measured at both 2/5 and 1/3 plateaus is shown. No excess noise is measured on the plateaus since no partitioning takes place and the reservoirs produce a noise free current.

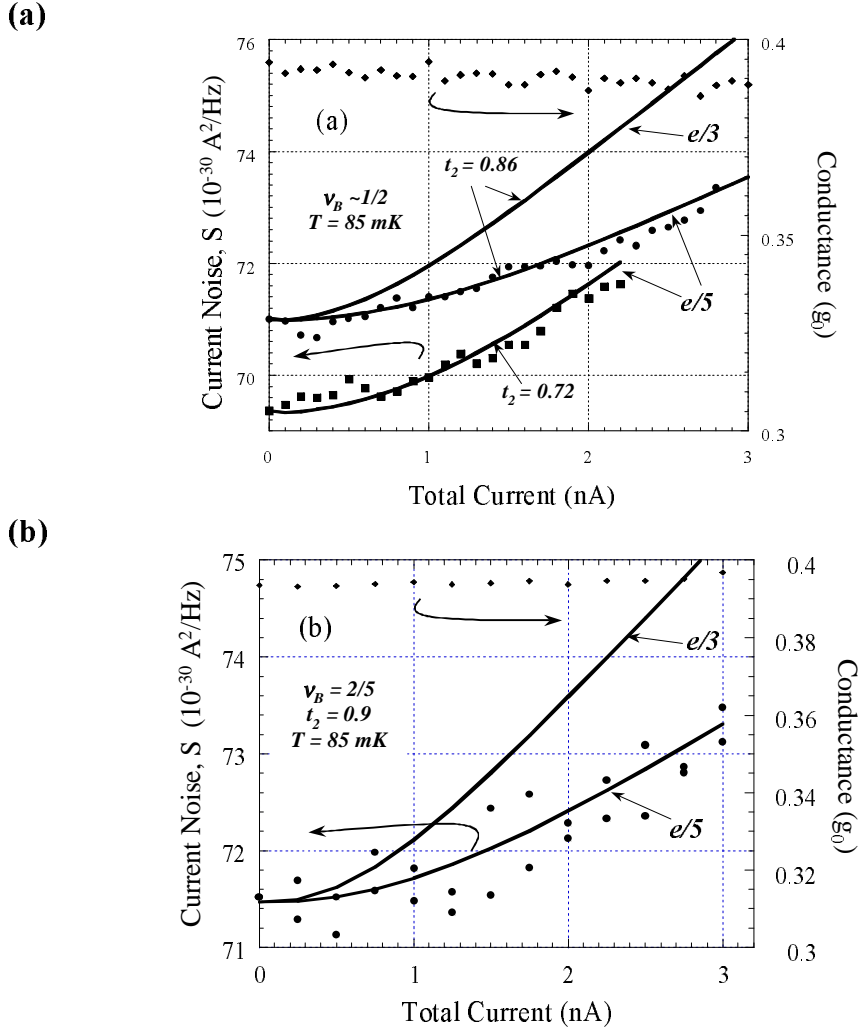

**Figure 3.** Measured noise of quasi-particles in the 2$^{nd}$ Composite Fermion channel. (a) Spectral density of current fluctuations against the total average current for transmission $t_2=0.86$ and $t_2=0.72$ at bulk filling factor 1/2 (solid circles, left scale). The 1$^{st}$ CF channel is fully transmitted and does not produce noise. Solid lines are given by Eq. (2) assuming charges $e/3$ and $e/5$ - as indicated. The sample's differential conductance, for transmission $t_2=0.86$ is also shown (solid dimonds, right scale). (b) Similar noise and conductance data for $t_2=0.9$ and bulk filling factor 2/5.

The current noise for weakly back-scattered quasi-particles in the second CF channel, at two different bulk fillings, is shown in Fig. 3. The first CF channel (the '*1/3*' channel) is fully transmitted and, in an independent channels model, does not produce noise (as seen in the inset of Fig. 2). Figure 3(a) shows the measured noise for two values of the transmission of the QPC, $t_2=0.86$ and $t_2=0.72$, at a bulk filling factor $v_B \sim 1/2$. Using a quasi-particle charge $q=e/5$ and transmission deduced from the average value of conductance (within the applied DC current range), the measured noise agrees well with the prediction of Eq. (2) throughout the whole range of driven DC current. Note that the differential conductance, and hence the deduced transmission, are rather constant in the full range of the measurement. For comparison, the expected noise for the same transmitted current but with the assumption of quasi-particles with a charge $q=e/3$ is shown for the case $t_2=0.86$ in the same figure. Similarly, Fig. 3(b) shows the noise data for a different bulk filling, $v_B=2/5$, and $t_2=0.9$. Even though the signal is weak, and thus scattering of the data points is relatively large, it clearly verifies that the charge of the quasi-particles is $e/5$. Note that there are no fitting parameters in these theoretical curves, as the transmission coefficient $t_2$ and the temperature of the electrons $T$ are both measured independently[4,17]. Here, again, we see a clear manifestation of transport in the 2$^{nd}$ CF channel with no contribution of the 1$^{st}$ channel, as one would expect from the CF model.

A similar agreement between the measured noise and Eq. (2) is obtained also with smaller transmission coefficients in the range $0.4 < t_2 < 0.6$, as is demonstrated by the example of $t_2=0.47$ in Fig. 4. The agreement of the measured noise with Eq. (2) for such a relatively large reflection coefficient is somewhat surprising. Theories[6,7,14] find a smooth variation of the measured charge of the quasi-particles, from $e/3$ for weak back scattering to $e$ for strong back scattering, in the first CF channel ($v=1/3 \rightarrow v=0$ at the QPC). Extending the same arguments to the transition $v=2/5 \rightarrow v=1/3$ at the QPC one would expect to measure a charge $e/5$ at large $t_2$ changing monotonically to a charge $e/3$ at small $t_2$ (see also ref. 14). We should note though that the differential conductance in this case is not entirely independent of the applied voltage. In general, non-linear I-V characteristic complicates the interpretation of our measurements, however, we expect that the modifications to Eq. (2), due to the energy dependence of $t$, are small. Aside from possible channel mixing the change in the conductance by a relatively small amount induces a maximal change in the thermal noise $4kT\delta g$ of some 25% of the measured excess noise. Also, the weak dependence of the expected shot noise on $t_2$ near $t=0.5$ (in a non-interacting Fermions picture) makes this non-linearity insignificant. In fact, we were unable thus far to probe the noise properties under even stronger reflection of the 2$^{nd}$ CF channel ($t_2<0.4$) because the *I-V* characteristic of our device becomes very non-linear at such low transmissions.

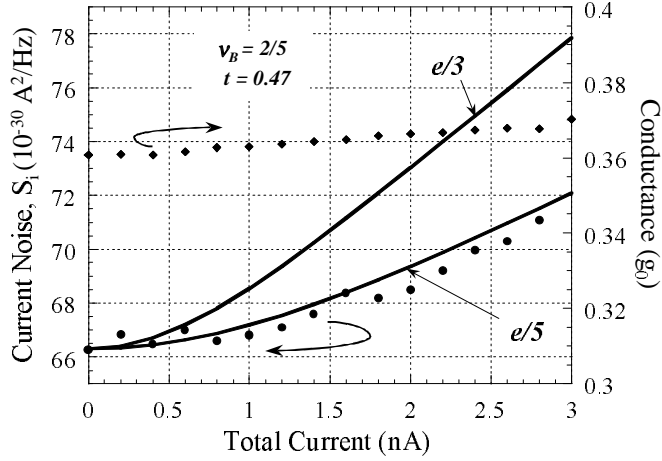

**Figure 4.** Measured Spectral density of current fluctuations against the total average current for relatively small transmission coefficient, $t_2=0.47$ (solid points, left scale). Solid lines are given by Eq. (2) assuming charges of $e/3$ and $e/5$ – as indicated. The sample's differential conductance is also shown (solid dimonds, right scale). The transmission $t_2$ is determined by the average differential conductance in the current range.

In conclusion, our shot noise measurements show that while the current of the 1$^{st}$ Composite Fermion (CF) channel in the FQH regime ($v=1/3$) is carried by quasi-particles with a charge $e/3$, the current in the 2$^{nd}$ CF channel ($v=2/5$) is carried by even smaller charges, $q=e/5$. The measurement demonstrated clearly that the two CF channels *live* independently with little mutual interactions – consistent with the CF model. The observation of a charge $q=e/5$ at filling factor $v=2/5$ is a demonstration that current-carrying charge can be different from the filling factor (or the fractional conductance). A surprising result, even though still preliminary, is that the charge remains $e/5$ when the backscattering of the 2$^{nd}$ CF channel is as strong as 50%; in contradiction with theories that predict a larger charge, approaching $e/3$ for a fully reflected 2$^{nd}$ CF channel.

## Acknowledgement

We would like to thank G. Bunin for manufacturing the transistors for the preamplifier and D. Mahalu for performing the e-beam lithography. The work was partly supported by a grant from the Israeli Science Foundation and by a grant from the Israeli Ministry of Science.